\PassOptionsToPackage{draft}{hyperref}
\documentclass[sigconf]{acmart}
\renewcommand\footnotetextcopyrightpermission[1]{} % removes footnote with conference information in first column

\usepackage{booktabs} 
\usepackage{paralist}
\usepackage{listings}
\usepackage[linesnumbered]{algorithm2e}

\usepackage{subcaption}

\newcommand{\indep}{\perp \!\!\! \perp}
\setcopyright{rightsretained}

\acmConference[SEAMS'20]{15th International Symposium on Software Engineering for Adaptive and Self-Managing Systems}{May 2020}{Seoul, South Korea}
\acmYear{2020}
\copyrightyear{2016}

\acmArticle{4}

\begin{document}
\title{Collective Risk Minimization via a Bayesian Model for Statistical Software Testing}

\author{Joachim Haensel, Christian M. Adriano, Johannes Dyck, Holger Giese}
\orcid{}
\email{joachim.haensel|christian.adriano|johannes.dyck|holger.giese @hpi.de}

\begin{abstract}
In the last four years, the number of distinct autonomous vehicles platforms deployed in the streets of California increased 6-fold, while the reported accidents increased 12-fold. This can become a trend with no signs of subsiding as it is fueled by a constant stream of innovations in hardware sensors and machine learning software. Meanwhile, if we expect the public and regulators to trust the autonomous vehicle platforms, we need to find better ways to solve the problem of adding technological complexity without increasing the risk of accidents. We studied this problem from the perspective of reliability engineering in which a given risk of an accident has severity and probability of occurring.  Timely information on accidents is important for engineers to anticipate and reuse previous failures to approximate the risk of accidents in a new city. However, this is challenging in the context of autonomous vehicles because of the sparse nature of data on the operational scenarios (driving trajectories in a new city). \emph{Our approach} was to mitigate data sparsity by reducing the state space through monitoring of multiple-vehicles operations. We then minimized the risk of accidents by determining proper allocation of tests for each equivalence class. \emph{Our contributions} comprise (1) a set of strategies to monitor the operational data of multiple autonomous vehicles, (2) a Bayesian model that estimates changes in the risk of accidents, and (3) a feedback control-loop that minimizes these risks by reallocating test effort. \emph{Our results} are promising in the sense that we were able to measure and control risk for a diversity of changes in the operational scenarios. We evaluated our models with data from two real cities with distinct traffic patterns and made the data available for the community.
\end{abstract}

\maketitle
% ============================================================================================
%
% ============================================================================================
\section{Introduction}\label{sec:intro}
\noindent
After a promising start at the DARPA competition~\cite{berger2012autonomous} and  extensive testing in city streets~\cite{wired_davis_alex_2017}, autonomous vehicles started to finally look reliable. This was a particularly ambitious outlook for a technology that is so reliant on a constant stream of innovations in blackbox machine learning models~\cite{burton2017making} and for which one cannot fully anticipate all of the operational scenarios for testing ~\cite{stewart_why_2018wired}. As a consequence, between the years of 2015 and 2019, the number of companies testing cars in the streets of California increased six-fold (from 11 to 65), while the number of accidents increased almost 12-fold (from nine to 104)~\cite{DMV_California}. Sadly, it was also within this period that the first fatal crashes started to happen~\cite{ntsb2019Uber,simonite2016Tesla}.

The engineering challenge is how to provide safety-critical assurances when the operational scenario changes~\cite{chechik2019software,bertolino2019changing}. The approach has been to design systems with self-adaptation capabilities~\cite{deLemos2017assurances}, for instance, feedback control-loops~\cite{cheng2009software} and runtime models~\cite{vogel2010adaptation}. These models provide a principled framework to design complex adaptive behaviors that allow systems to handle unanticipated changes in their operational environments. This also implied that the testing of these systems happens in the presence of adaptions and the uncertainties in the models that generated them~\cite{Giese:2014ca, esfahani2013uncertainty, ramirez2012taxonomy}.
 
To mitigate model uncertainties, different approaches were proposed for testing self-adaptive systems (SAS)~\cite{siqueira2016characterisation}, for instance, robustness testing~\cite{camara2014testing,hansel2015testing}, online testing~\cite{hielscher2008framework}, runtime-based assurance techniques~\cite{cheng2014using}, and risk-based testing~\cite{reichstaller2018risk}\cite{matus2018patent}.

However, the current state of the art still lacks in terms of guidance to allocate tests when the test evidence is partial, and the input data is sparse. These are characteristic of the unanticipated scenarios faced by a SAS operating in a changing environment ~\cite{bertolino2019changing,chechik2019software} and they stem from the fact that tests cannot guarantee the absence of failures~\cite{dijkstra1970notes} because there is no number of tests that can uncover all defects in a software.

Hence, we approach test allocation from the perspective of software reliability testing \cite{brown_testing_1975}, which allows estimating the reliability of software even when there are no identified failures. \cite{miller_estimating_1992}. The approach is based on  detecting the failures that might manifest more frequently according to an operational profile \cite{Musa1993}. Intuitively, it consists of allocating the tests in a way that mirrors how the software might be executed by the end-user \cite{bertolino2007testing}.

\emph{The problem} that we focus on is how to find a test allocation strategy that minimizes the risk of accidents in a new environment with sparse data. We partition the problem in two research questions: (1) how to estimate the risks of accidents before moving to a new environment, and (2) how well can we mitigate these risks by properly allocating tests. 

\emph{Our Approach} was based on mitigating data sparsity by reducing the state space. For that, we systematically identified and monitored equivalence classes of multiple-vehicles operations. We used these classes to allocate tests in a proportion that minimizes the risk of accidents. To discover these test allocations, we applied statistical test methods~\cite{GardinerK1999} that measure the risk as a function of hazard (severity) of a failure and the corresponding reliability of each equivalence class. By "reliability", we mean "the probability of failure-free operation in a specified environment over a defined period of time" \cite{musa1987engineering}. This definition follows the DIN400-41 standard~\cite{din40041} and is widely adopted in the automotive software engineering practice~\cite{zurawka2016automotive}. 

\begin{figure}
\centering
	\includegraphics[width=0.49\textwidth]{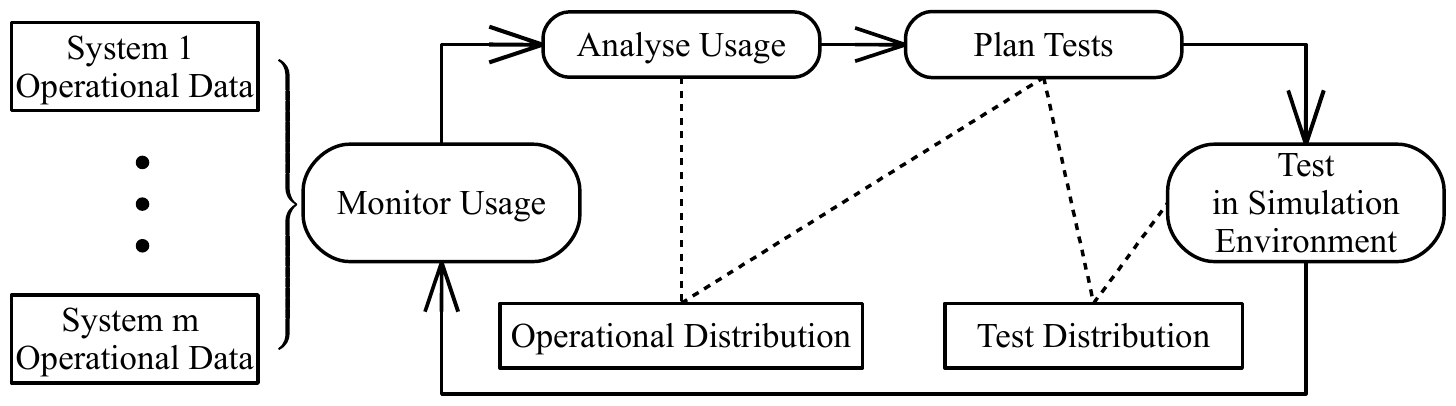}
	\caption{General methodology}
	\label{fig:workflow}
\end{figure}

\emph{Our contribution} is a general four-step methodology (Figure \ref{fig:workflow}) that works as a template to instantiate different self-adaptive strategies (section \ref{sec:approach}) and comprises two methods: 
\begin{enumerate}
 \item a monitoring method to mitigate data sparsity by collecting operational data of multiple autonomous vehicles (\textit{System n Operational Data} and \textit{Monitor Usage} step)
 \item a statistical method to estimate risks (\textit{Analyse Usage} and \textit{Plan Tests} based on the \textit{Operational and Test Distribution})
\end{enumerate} In order to actively pursue risk-based goals, these methods operate in a feedback control-loop that updates the operational and test distribution and executes the tests to check the system for failures.

\emph{Our results} are promising in a sense that we were able to measure and control risk for a diversity of operational scenarios. To allow the reproduction of our results, we made the procedures, models, and data publicly available to the community.

We structured the paper as follows. In Section \ref{sec:examplearch} we describe the example scenario and corresponding architecture. In Section \ref{sec:prelim} we explain the preliminaries that are the foundations of our approach. Our approach is then detailed in Section \ref{sec:approach}, followed by the solution of optimization problems in Section \ref{sec:optimization}, evaluated in Section \ref{sec:evaluation}, and discussed in Section \ref{sec:discussion}. In Section \ref{sec:related} we position our contributions with respect to other related work. Finally, in Section \ref{sec:conclusion}, we summarize our contributions and future work.

% ============================================================================================
%
% ============================================================================================
\section{Example Scenario and Architecture}
\label{sec:examplearch}
\noindent
The scenario and corresponding architecture introduce the basic intuitions about the actors and the objects involved in the self-adaptation of the autonomous vehicle's operations. These definitions will later be used to derive the statistical models and to plan their empirical evaluations.  
\subsection{Architecture}
\begin{figure}
  \begin{subfigure}[c]{0.21\textwidth}
    \includegraphics[width=\textwidth]{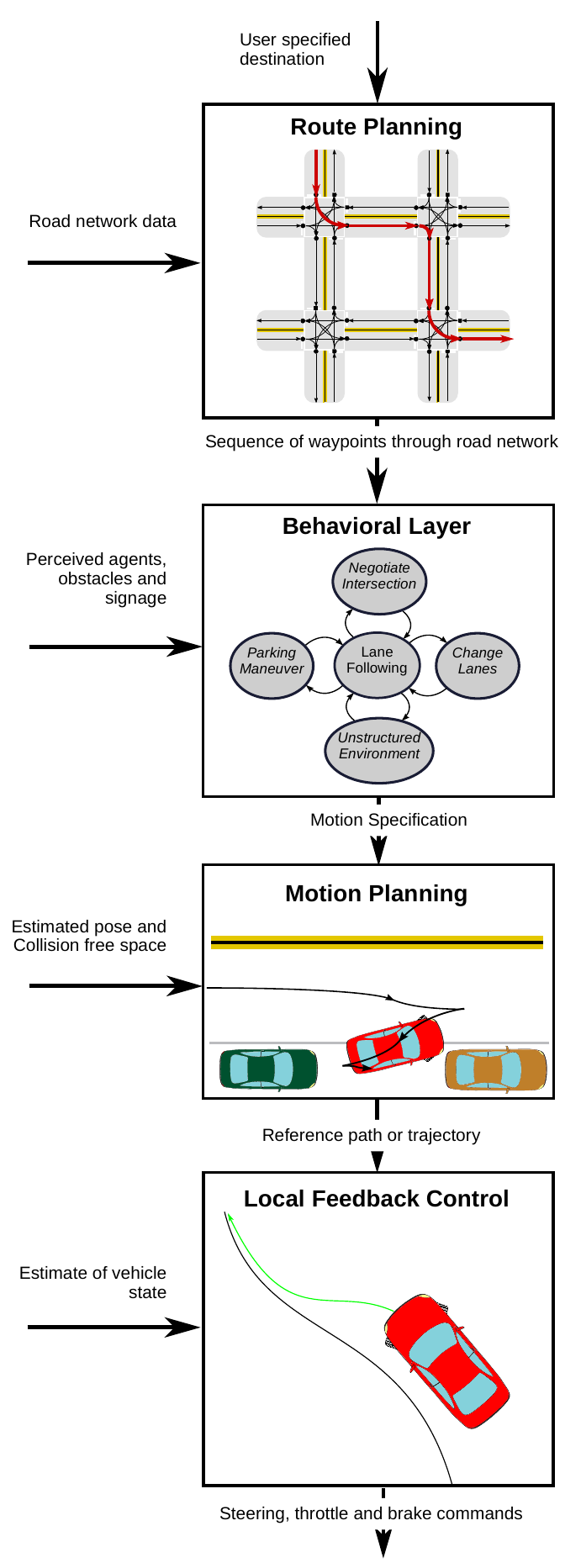}
  \end{subfigure}
  \hfill
  \begin{subfigure}[c]{0.21\textwidth}
    \includegraphics[width=\textwidth]{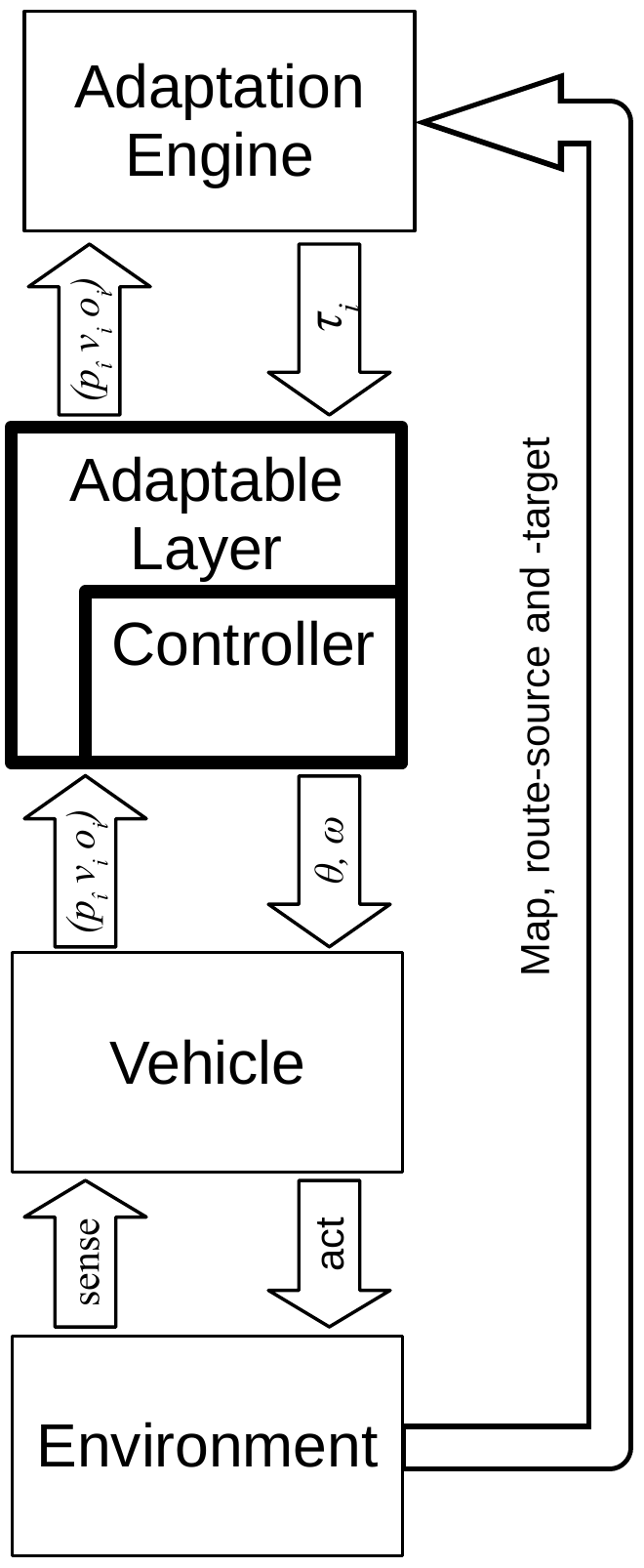}
  \end{subfigure}
  \caption{Decision-making hierarchy used in self-driving cars according to \cite{Paden2016} and the self-adaptation focused architecture}
  \label{fig:vehiclearch}
\end{figure}

\noindent
In our scenario, self-adaptive systems will be represented by autonomous vehicles. The software architecture for the vehicle is a version of the decision-making hierarchy described in \cite{Paden2016} (Figure \ref{fig:vehiclearch}, left). Since we focus on self-adaptive systems, we transferred the decision-making hierarchy into a two-layered architecture (Figure \ref{fig:vehiclearch}, right) with an adaptation engine (route planning, behavioral layer, motion planning) and an adaptable layer (local feedback control). 
 
\noindent
The scenario consists of tasks associated with the vehicles and with target destinations for an area defined by a map. Based on this information, the adaptation engine computes the feasible route and the velocities along this route. 

\begin{figure}
	\includegraphics[width=0.48\textwidth]{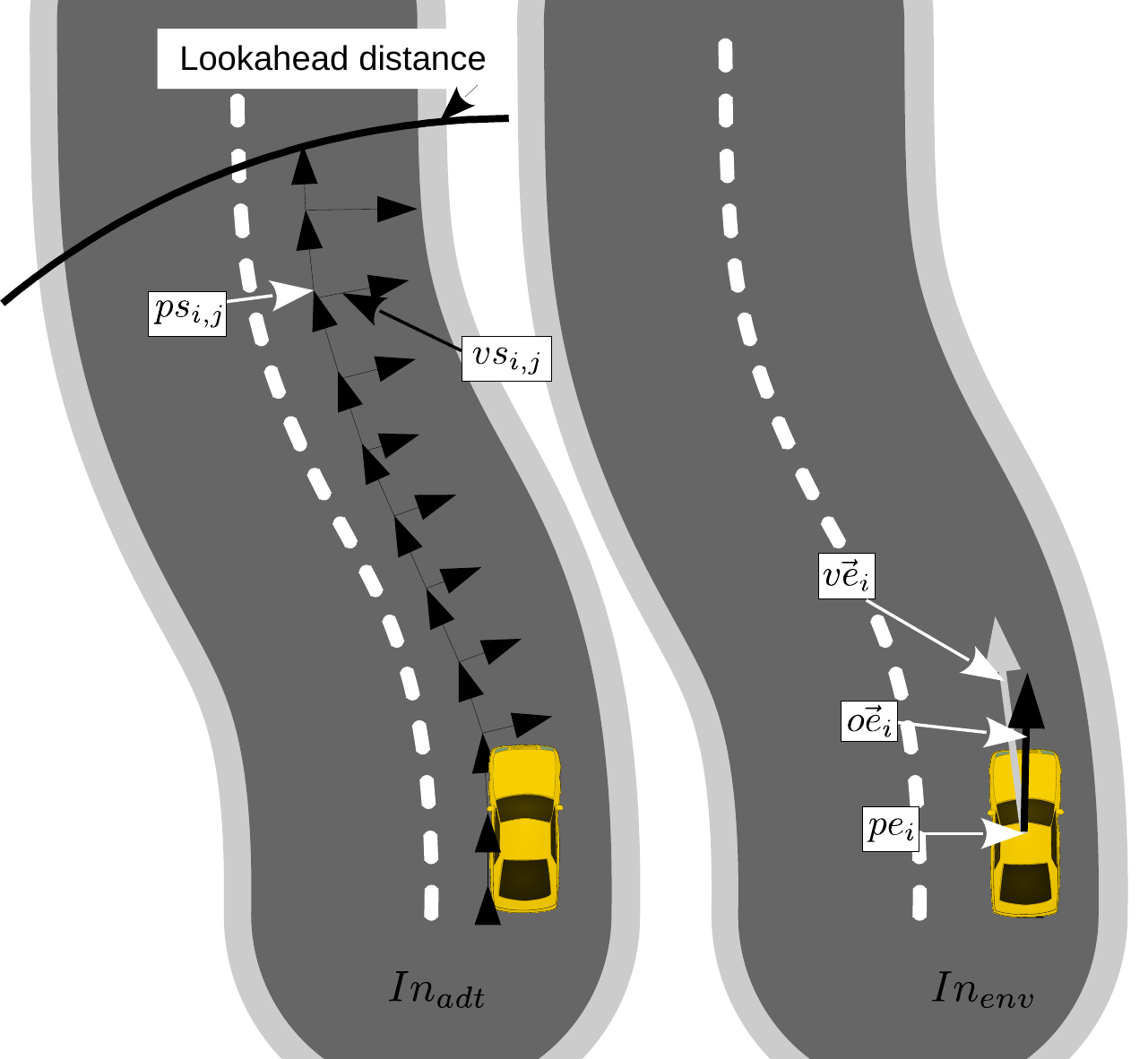}
	\caption{Adaptation engine and environment input}
	\label{fig:controlled_input}
\end{figure}

\noindent
During the car ride, the adaptation layer provides a fixed-length planning horizon from the pre-computed route. This planning horizon $\tau$ contains a set of directions and a set of velocities along this path (Figure \ref{fig:controlled_input}, left). The adaptable layer uses this information to compute the steering angle (directions) and the velocity. Whenever the vehicle passes the first vector of the planning horizon, a new horizon is provided by the adaptable layer. 

\noindent
The adaptable layer senses the current location ($p$), velocity ($\vec{v}$), and orientation ($\vec{o}$) from the environment (Figure \ref{fig:controlled_input}, right). This information is reported back to the adaptation layer and used there as a source for the next planning horizon. It could, for example, reflect this information when necessary if the vehicle is too far from a planned position. 

\noindent
The central part of the adaptable layer is the control algorithm, which computes the set of values for a steering angle ($\theta$) of an Ackermann-steering and the wheel rotation speed ($\omega$) of the front wheels. The computation is based on the provided path $\tau$ from the planning horizon and the corresponding velocities in addition to $p$, $\vec{v}$, and $\vec{o}$ from the environment. 

\noindent
We define the input for the adaptable layer from the adaptation layer as follows (see Figure \ref{fig:controlled_input}, left): 
$In_{adt} = \{\tau_0, .., \tau_n\}$ with $\tau_i = \{ts_0, .., ts_m\}$ the $i$th planning horizon for the adaptable layer. An element in a planning horizon $\tau_i$ is defined by $ts_j = (ps_j, vs_j)$ with $ps_j$ as a point on the street to be passed and $vs_j$ the set-velocity for this point. Elements have a fixed distance $d$ to each other, so that: $\forall j: |ps_j - ps_{j-1}| = d$. The input from the environment to the adaptable layer and to the adaptation engine is defined in this way (Figure \ref{fig:controlled_input}, right): $In_{env} = {e_0, \ldots, e_n}$ with $e_i = (\vec{pe_i}, \vec{ve_i}, \vec{oe_i})$, $\vec{pe_i}$ the current position, $\vec{ve_i}$ the current velocity and $\vec{oe_i}$ the current orientation of the vehicle. 

\noindent
In our approach, we target the adaptable layer with testing. The input space for the operational profile is, therefore, the input from the adaptation layer $In_{adt}$ and the environment $In_{env}$.

\subsection{Scenario}
\noindent 
Our scenario starts when the first risk estimations of the vehicle are carried out, which happens at system development-time. If this estimated risk is below a defined upper bound, the system will be released for deployment. After deployment, the adaptable layer is not in itself changed anymore. New software versions will require a new run of our approach. 

The vehicles will first be deployed in the city from which the risk assessment team assumed an operational profile for release testing. An operational profile consists of a discrete probability density function over the binned input space (\(In_{adt} \cap In_{env}\)). Intuitively, this function describes the likelihood that any set of inputs within a bin will be selected when the autonomous vehicle is used in the city streets.

\noindent
While driving, data from all deployed vehicles is transferred to the manufacturer, where the previously estimated operational profile is updated. Alongside the operational profile, the risk estimates will be updated as well. If necessary, additional tests will be allocated to lower the risk.  

\noindent
After a while, vehicles start to get deployed in new and yet unknown cities. Up to this point, the operational profile converged only to the operational profile of the original location of deployment. In the new deployment environment, it is expected that the current profile will have to be updated. This change in the operational profile will entail further re-estimations of risk and possibly further tests. In our scenario we will consider different strategies for deciding when and how many additional tests might be necessary.      
% ============================================================================================
%
% ============================================================================================
\section{Preliminaries}\label{sec:prelim}
\noindent
In this section, preliminaries are briefly summarized, and we establish some basic terminology.
\subsection{Software Reliability Testing}
\noindent
Software reliability~\cite{lyu2007software} and the more traditional term of reliability connected to hardware have been recognized as something very different already starting from 1970. In \cite{brown_testing_1975}, for example, software reliability is characterized by the relation between failures that always exist in software and the usage of software according to an operational profile. Unsurprisingly the partitioning of an input space (the operational profile as subdomains) together with a probability distribution for the profile is necessary for test allocation. Software Reliability Testing is also known as Statistical Software Testing.

Based on the 2016 update of the IEEE 1633 standard \cite{1633IEEEReliability}, we defined our process for software reliability testing in four activities: (1) build a set of statistical models to predict the reliability risk, (2) update the models to reflect changes in the operational environment, (3) perform sensitivity analysis to identify the subdomains in the operational profile that are more sensitive to change, and (4), based on that, allocate additional tests to minimize increases in risk. 

\noindent
Most of the software reliability testing techniques rely on a history of test failures (e.g., growth models) \cite{malaiya2002software}\cite{amin2013approach}, usage information (operational profile) \cite{miller_estimating_1992}, or a combination of the both \cite{xiao2018optimal}. Since we aim to estimate risk even in the absence of failures, we do not rely on a history of failures. Conversely, our models are based solely on the distribution of usage and tests across the operational profile subdomains.

We termed the software inputs as \textit{demands}, which have an associated  probability of failing for a demand (\textbf{p}robability of \textbf{f}ailure on \textbf{d}emand or short \(pfd\), \cite{littlewoodbev_validation_1993}). In order to compute \(pfd\), a software is tested with inputs according to the operational profile. Even if these tests do not reveal failures, an estimate of \(pfd\) can be made based on a Bayesian model \cite{miller_estimating_1992}.

\noindent
The idea is to get an estimate of \(pfd\) and combine that with the operational profile. The operational profile is denoted by a discrete probability distribution \(P = \{p_1, \ldots, p_n\}\) over the subdomains (or bins in statistical terminology). In our example, the subdomains are derived over the input space of \(In_{adt} \cap In_{env}\). \(pfd\) is computed with the Laplace Rule of Succession: \(pfd = \frac{1}{2+t}\), where t denotes the number of successful tests. When applied to the subdomains, the overall reliability estimate becomes: \(pfd = \sum^n_i p_i \frac{1}{2+t_i}\) with \(t_i\) the number of tests applied to each subdomain.   
% ============================================================================================
%
% ============================================================================================
\subsection{Statistical Software Testing for Risk Analysis}\label{subsec:sst4ra}
\noindent
 As outlined in detail by Gardiner in \cite{GardinerK1999}, SST can also be employed for risk analysis to help estimating risk. Therefore, it is suggested to identify \emph{demands} as specifically critical scenarios such as, for example, the occurrences of a \emph{tire blowout} as a demand $e$ and then estimate the likelihood of such a demand by $\lambda_e$, the severity of an accident in case of such a demand as $\epsilon_e$, and estimate the likelihood of the occurrence of the demand $e$ employing statistical software testing employing a dedicated environment simulation by $\mathop{pfd}_e = 1/(2 + t_e)$ for $t_e$ the number of tests assuming that all tests where showing no accident.
As outlined by Gardiner in \cite[p.~164]{GardinerK1999}, statistical software testing without failures can also be employed for risk analysis. It is suggested to identify a particular critical scenario $e$ and the likelihood of this critical scenario $\lambda_e$. We also associate a value $\epsilon_e$, which reflects the hazard caused by $e$ in case the system fails for this critical scenario. Based on our example, a tire-blowout could be a situation for which the autonomous vehicle is required to react in a way that does not cause harm.

\noindent
Given a set \(H\) of all critical scenarios,
\begin{equation}\label{eq:ra-std}
    risk(per\ demand) 
  = 
    \sum_{e \in H} {\mathop{pfd}}_e \lambda_e \epsilon_e
\end{equation}
results in the estimate of the current risk. All tests for this estimation are carried out in an environment where critical scenarios can be simulated. Because the analysis of risk requires the inclusion of hazard \(\epsilon_e\), we no longer depend only on the occurrence distribution of \(\lambda\), but also on the severity distribution \(\epsilon\). On the other hand, any operational profile that is orthogonal to the occurrence of a hazard scenario will be omitted to reflect the assumption of a steady-state required for this approach of testing. 
% ============================================================================================
%
% ============================================================================================
\section{Approach}\label{sec:approach}
\noindent
Our approach combines Bayesian modeling of statistical software testing for risk analysis with the monitoring of an operational profile, as detailed in Subsection \ref{subsec:approach-sst4ra4op}. In Subsection \ref{subsec:approach-od@rt}, we describe how we update our prior knowledge of the operational profile that reflects a change in the environment. In Subsection \ref{subsec:approach-sst4ra@rt}, we explain how updates in the profile affect the estimates of upper-bound risk. We present three different strategies that either (1) keep risk at the same level with possibly infinite additional tests, (2) lower the risk continuously by testing with a fixed number of tests, (3) or combine both approaches (see Figure \ref{fig:strategy}).
% ============================================================================================
\subsection{Statistical Software Testing for Risk Analysis for Operational Profiles}\label{subsec:approach-sst4ra4op}
\noindent
To employ statistical software testing for risk analysis \cite{GardinerK1999}, as introduced in Section~\ref{subsec:sst4ra}, we adjusted the setting to avoid the unreasonable strong assumption of a steady-state of the system. Instead, the operational profiles $p_1, \dots, p_n$ over the equivalence classes (subdomains) $I$ with $|I|=n$ are used to allocate the test effort so that the occurrences of a demand are tested for each of the equivalence classes. This can be done by adjusting Equation \ref{eq:ra-std} accordingly.
\begin{equation}\label{eq:ra-apprach}
    risk(per\ demand) 
  = 
    \sum_{e \in H} 
      \sum_{i \in I} 
        {\mathop{pfd}}_{ei} \lambda_e \epsilon_e
\end{equation}
where $pfd_{ei}$ is obtained as
%\begin{equation}\label{eq:pdf-apprach}
$ 
     {\mathop{pfd}}_{ei} 
   = 
     (1/(2+t_{ei}))
     p_i
$%

\noindent
If during development-time a required upper bound $UB$ for an estimated operational profile $p_1, \dots, p_n$ has to be ensured, we thus will consider
\begin{equation}\label{eq:ra-apprach-offline}
    \sum_{e \in H} 
      \sum_{i \in I} 
        \frac{1}{2+t_{ei}} p_i \lambda_e \epsilon_e
  \leq
    UB
,
\end{equation}
where we assume that $p_i \indep$ \(\lambda_e\) (independence). Moreover, $t_{ei}$ have to be found such that Equation~\ref{eq:ra-apprach-offline} holds while the cost $\sum_{e \in H} \sum_{i \in I} t_{ei}$ is minimized.
Alternatively, if sufficient resources for testing are available such that $m$ tests with $m > \sum_{e \in H} \sum_{i \in I} t_{ei}$ can be done, it would be more appropriate to actually minimize the outcome of Equation~\ref{eq:ra-apprach-offline} by taking the cost limit $m$ into account.

% ============================================================================================
\subsection{Operational Distributions at Run-Time}\label{subsec:approach-od@rt}
\noindent
The distribution of an operational profile might not remain the same after deployment~\cite{Musa1993}. For this reason, the reliability engineering practice recommends updating the operational distribution after deployment~\cite{1633IEEEReliability}. This is in line with the need to update the risk estimates when the environment changes. For a single system, this would result in monitoring the current state of the system by counting inputs for each of the operational profiles subdomains (equivalence classes, bins).

\noindent
The current state of the operational profile is derived from the occurrence counts \(O = \{o_1, \ldots, o_n\}\), where for each subdomain (with \(|X| = \sum^n_1 x_i, X = \{x_1, \ldots x_n\}\)):
\begin{equation}\label{eq:dist-offline}
	P = \{{\frac{o_1}{|O|}, \ldots, \frac{o_n}{|O|}}\} = \{p_1, \ldots, p_n\}
\end{equation}
The monitoring of updates \(U = \{u_o, \ldots,u_n\}\) will simply be added to the current occurrence counts, which will generate an updated profile
\begin{equation}\label{eq:dist-at-runtime}
	P' = \{{\frac{o_1 + u_1}{|O| + |U|}, \ldots, \frac{o_n + u_n}{|O| + |U|}}\} = \{p'_1, \ldots, p'_n\}
\end{equation}
In the case of autonomous vehicles, we face the challenge of an input space that is very large. This happens even if the subdomains had been selected in an optimal way. As a consequence, the profile would be updated slowly. However, autonomous vehicles are not single instance systems. Instead, they are deployed on a large-scale basis. This allows us to collect monitoring input from all running system instances and update a centralized profile, similarly as it is done with the single instance (Equation \ref{eq:dist-at-runtime}). The benefit of updating from multiple instances is the faster synchronization of the profile with the reality of usage.

% ============================================================================================
\subsection{Statistical Software Testing for Risk Analysis at Run-Time}\label{subsec:approach-sst4ra@rt}
\begin{figure}
\centering
	\includegraphics[width=0.45\textwidth]{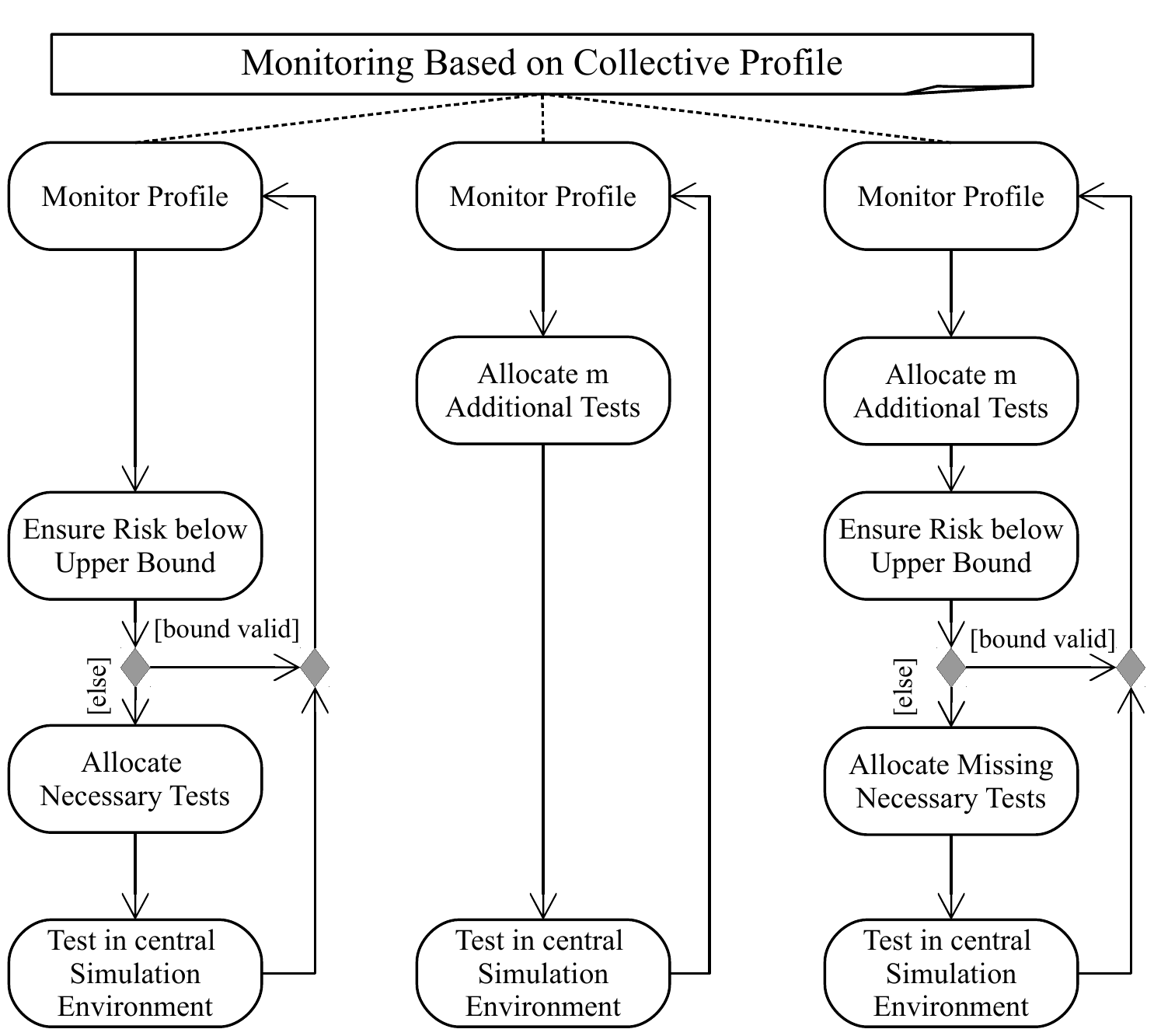}
	\caption{Feedback control-loop strategies for (1) risk maintenance, (2) risk improvement and (3) combination of risk improvement and maintenance}
	\label{fig:strategy}
\end{figure}
\noindent
The extension introduced in Subsection \ref{subsec:approach-sst4ra4op} implies that the risk dependency goes beyond the hazard scenario likelihood \(\lambda_e\) and the severity \(\epsilon_e\). The risk also depends on the impact of changes in the operational profile.

Consequently, any change in the operational profile at run-time will possibly result in a reduced accuracy of the risk that was assessed during system development-time.

\noindent
Therefore, we propose to use the develop-time risk assessment as a risk baseline, which we inherit and work to minimize at run-time. This results in three strategies:
(1) maintenance of a required upper bound for the risk (Figure \ref{fig:strategy} left) and 
(2) steering the testing efforts while the system is operating, such that the lowest possible upper bound for the risk can be established (Figure \ref{fig:strategy} middle and right).
(3) combination of (1) and (2) (Figure \ref{fig:strategy} right).

\subsubsection{Maintaining Required Upper Bounds for Risk at Run-Time}
\noindent
Given an updated operational profile $P'$ we have to ensure that
\begin{equation}\label{eq:ra-apprach-maintain}
    \sum_{e \in H} 
      \sum_{i \in I} 
        \frac{1}{2+t'_{ei}} p'_i \lambda_e \epsilon_e
  \leq
    UB
,
\end{equation}
\noindent
with $t'_{ei} \geq t_{ei}$ by running $\sum_{e \in H} \sum_{i \in I} t'_{ei} - t_{ei}$ additional tests while the system is operating.
Equation~\ref{eq:ra-apprach-maintain} can at first be used to check whether there is at all the need to do additional tests (see "\texttt{computeRisk(op',tests) > UpperBound}" from Algorithm \ref{alg:strat1}). If the operational distribution evolves towards the direction that the statistical testing could attain the required upper bound for the risk with fewer tests, then we do not need additional tests. Otherwise, it would be necessary to do additional tests to ensure that the upper bound still holds. This implies finding $t'_{ei}$ such that Equation~\ref{eq:ra-apprach-maintain} holds while the cost $\sum_{e \in H} \sum_{i \in I} t'_{ei} - t_{ei}$ is minimized  (see "\texttt{computeReqAddTests(op',tests)}" from Algorithm \ref{alg:strat1}).
\begin{algorithm}[]
\SetAlgoLined
 op' := monitorOperationalProfile()\;
 \If{computeRisk(op',tests) > UpperBound}{
  tests' := computeReqAddTests(op',tests)\;
  executeAddTest(tests')\;
  op := op'\;
  tests := tests $\cup$ tests'\;
 }
 \caption{Strategy 1, maintaining upper bound}
 \label{alg:strat1}
\end{algorithm}
\noindent
\subsubsection{Minimize Upper Bounds for Risk at Run-Time}
\noindent
Assuming fixed resources for testing while the system is running, allows to do $m$ additional tests. Hence, the task becomes to minimize risk for the operational profile that have evolved from $P$ to $P'$ 
\begin{equation}\label{eq:ra-apprach-minimize}
\min
    \sum_{e \in H} 
      \sum_{i \in I} 
        \frac{1}{2+\hat{t}_{ei}} p'_i \lambda_e \epsilon_e
,
\end{equation}
\noindent
respecting that the additional costs $\sum_{e \in H} \sum_{i \in I} \hat{t}_{ei} - t_{ei}$ must be bound to $m$ (see "\texttt{computeOptAddTests(op',tests,m);}" in Algorithm \ref{alg:strat2}).
\begin{algorithm}
op' := monitorOperationalProfile()\;
tests' := computeOptAddTests(op', tests, m)\;
executeAddTest(tests')\;
op := op'\;
tests := tests $\cup$ tests'\;
\caption{Strategy 2, minimize risk with additional tests}
\label{alg:strat2}
\end{algorithm}

\subsubsection{Maintaining and Minimizing Risk at Run-Time}
\noindent
Minimizing Equation~\ref{eq:ra-apprach-minimize} will not always guarantee that Equation~\ref{eq:ra-apprach-maintain} still holds for an upper bound $UB$. In this case, we suggest a combination of both strategies. Here, minimization can only be considered if for all $e \in H$ and $i \in I$ we have $\hat{t}_{ei} \geq t'_{ei}$ (see Algorithm \ref{alg:strat3}).
\begin{algorithm}
op' := monitorOperationalProfile()\;
tests'' := computeOptAddTests(op', tests, m)\;
tests := tests $\cup$ tests'\;
\If{computeRisk(op',tests) > UpperBound}{
  tests'' := computeReqAddTests(op',tests)\;
  executeAddTest(tests'' $\cup$ tests')\;
  tests := tests $\cup$ tests''\;
}
op := op';
\caption{Strategy 3, combined strategy}
\label{alg:strat3}
\end{algorithm}

% ============================================================================================
%
% ============================================================================================
\section{Upper Bound Minimization or Test Resource Optimization}\label{sec:optimization}
\noindent
We have established two optimization problems relevant to our approach:
\begin{inparaenum}
\item Given an upper bound $\mathit{UB}$ on risk per demand, minimize the sum of tests so that risk per demand lies beneath the upper bound (minimize $\sum_{e \in H}\sum_{i \in I} t_{ie}$ in Equation~\ref{eq:ra-apprach-offline}) and
\item given a fixed sum of tests, find a distribution of tests to minimize risk per demand (minimize $risk(per demand)$ in Equation~\ref{eq:ra-apprach}).
\end{inparaenum} 

A general solution to these minimization problems then applies to offline risk analysis as described in Section~\ref{subsec:approach-sst4ra4op}. From this we also derive solutions for the strategies applied at runtime (Section~\ref{subsec:approach-sst4ra@rt}). 

\subsection{Minimizing Tests Given an Upper Bound on Risk}
\noindent
Solving this optimization problem requires minimizing the objective function
\begin{equation*}
T = \sum_{e \in H} \sum_{i \in I} t_{ie} \tag{sum of tests}
\end{equation*}
for $t_{ie} \geq 0$ and under the inequality constraint
\begin{equation*}
\sum_{e \in H} \left(\lambda_e \epsilon_e \sum_{i \in I} \frac{p_i}{2 + t_{ie}}\right) \leq \mathit{UB} \tag{risk per demand}
\end{equation*}
given an upper bound $\mathit{UB}$ on risk per demand. We can find a real-valued solution by constructing the problem generalized as a Lagrange function and using the Karush-Kuhn-Tucker (KKT) conditions. Since both the objective function and the inequality constraint (for $t_{ie} > -2$) are convex, checking satisfaction of the KKT conditions is fairly straightforward, and the solution is indeed a global minimum (again, given $t_{ie} > -2$). We get
\begin{equation}
\label{eq:ra-solution-minimize-T-t}
t_{ie} = \frac{\sqrt{\lambda_e \epsilon_e p_i}}{\mathit{UB}} \left(\sum_{j \in H} \sqrt{\lambda_j \epsilon_j}\right) \left(\sum_{k \in I} \sqrt{p_k}\right)  -2
\end{equation}
and a corresponding lower bound on the number of tests
\begin{equation}
\label{eq:ra-solution-minimize-T-T}
T \geq \frac{1}{UB}\left(\sum_{e \in H} \sqrt{\lambda_e \epsilon_e}\right)^2 \left(\sum_{i \in I} \sqrt{p_i}\right)^2 - 2\left|H\right|\left|I\right|\text{.}
\end{equation}

However, we need integer test numbers -- and this solution is real-valued. Also, while Equation~\ref{eq:ra-solution-minimize-T-t} ensures that values $t_{ie}$ will be greater than $-2$, they can be negative. Although this will only happen if $\mathit{UB}$ is large in relation 
%\todo{evaluation?} 
to values $\lambda_e$, $\epsilon_e$, $p_i$, and the sums of Equation~\ref{eq:ra-solution-minimize-T-t}.%, it is possible.

We propose to solve both problems as follows: after determining the optimal real-valued solution, we go through all values $t_{ie}$ in an iterative fashion, rounding up or down with a minimum of $0$ as needed. When values are rounded up (which is the default), the corresponding decrease in risk is accumulated. Whenever the increase in risk from rounding down a value would be lower than the accumulated decrease in risk, we will round the value down instead and subtract the increase in risk from the accumulated decrease. 

We could also choose the naive solution of rounding up all values $t_{ie}$ to the next integer, with a minimum of $0$. This will lead to a worse result with respect to the number of tests; however, simply rounding up all values can be done while computing the real-valued solution, and all values $t_{ie}$ can be computed in parallel if desired. For the strategy above, dependencies between values are likely to make parallel execution more challenging and less effective.

\subsection{Minimizing Risk Given a Fixed Sum of Tests}
\noindent
To solve this optimization problem, we switch the objective function and constraint from before. We need to minimize
\begin{equation*}
R = \sum_{e \in H} \left(\lambda_e \epsilon_e \sum_{i \in I} \frac{p_i}{2 + t_{ie}}\right) \tag{risk per demand}
\end{equation*}
for $t_{ie} \geq 0$ and under the equality constraint
\begin{equation*}
T = \sum_{e \in H} \sum_{i \in I} t_{ie} \tag{sum of tests}
\end{equation*}
given a sum of tests $T$. Since we have an equality constraint instead of an inequality constraint, this problem is easier to solve. We do not need the KKT conditions, only the problem's Lagrange function and its partial derivatives. Again, the functions are convex for $t_{ie} > -2$. We get
\begin{equation}
\label{eq:ra-solution-minimize-R-t}
t_{ie} = \frac{\sqrt{\lambda_e \epsilon_e p_i}(T + 2 \left|H\right|\left|I\right|)}{\left(\sum_{j \in H} \sqrt{\lambda_j \epsilon_j}\right) \left(\sum_{k \in I} \sqrt{p_k}\right)} - 2
\end{equation}
and a resulting lower bound on risk per demand of
\begin{equation}
\label{eq:ra-solution-minimize-R-R}
\mathit{R} \geq \frac{\left(\sum_{e \in H} \sqrt{\lambda_e \epsilon_e}\right)^2 \left(\sum_{i \in I} \sqrt{p_i}\right)^2}{T + 2\left|H\right|\left|I\right|}\text{,}
\end{equation}
which, unsurprisingly, is equivalent to a rearranged Equation~\ref{eq:ra-solution-minimize-T-T} when $R$ is substituted for $\mathit{UB}$.

As before, this is a real-valued solution, and values $t_{ie}$ may lie between $-2$ and $0$ for a comparatively low number of tests $T$. Also, we cannot simply round up values $t_{ie}$ because their sum might then exceed $T$. Instead, we can round down values, starting with the lowest value $t_{ie}$, and accumulate the sum of values thusly subtracted (by rounding). Whenever the sum exceeds or equals the difference required to round up the next value $t_{ie}$, we round up instead and subtract the difference from the accumulated sum. Negative values are set to $0$ -- and if the accumulated sum falls below $0$, this can be compensated with the next value(s). This process leaves the sum $T$ of tests unchanged. While it does not necessarily result in the optimal (integer) solution with respect to risk, it brings us reasonably close (cf. Section~\ref{sec:evaluation}) to the lower bound on risk per demand (Equation~\ref{eq:ra-solution-minimize-R-R}).
\subsection{Optimization for Profile Change}\label{subsec:optprofilechange}
\noindent
For risk analysis at run-time, we need extensions of the previous optimization problems, in order to incorporate the knowledge of pre-existing tests. The computation of required additional tests (strategy 1) needs a minimization for a given upper bound and pre-existing tests. A steady number of additional tests per time (strategy 2) asks for a minimization of risk given a number of tests and pre-existing tests. Finally, a steady number of additional tests together with more tests when needed (strategy 3) is covered by a combination of the solutions of the other optimization problems.    
\subsubsection{Minimizing Tests}
\noindent
The objective is to find the minimal number of additional tests required if an upper bound should be kept, based on an existing distribution of tests. Since we do not need additional tests for bins (subdomains) where the probability has dropped, we first compute the risk value for all those bins. Given $I' = \{i \mid p'_i \leq p_i\}$ as the set of the respective indices, we have 
\begin{equation}
R_{cov} = \sum_{e \in H} \left(\lambda_e \epsilon_e \sum_{i \in I'} \frac{p'_i}{2+t_{ie}} \right). %\forall i | p'_i \leq p_i
\end{equation}

\noindent
For the remaining bins, where the new \(p'_i\) are more likely than the old \(p_i\), we first compute new intermediate \(t^*_{ie}\) for $i \in I \setminus I'$, taking the calculated risk value $R_{cov}$ into account:
\begin{equation}
\label{eq:ra-solution-minimize-update}
t^*_{ie} = \frac{\sqrt{\lambda_e \epsilon_e p'_i}}{\mathit{UB - R_{cov}}} \left(\sum_{j \in H} \sqrt{\lambda_j \epsilon_j}\right) \left(\sum_{k \in I \setminus I'} \sqrt{p'_k}\right)  -2%, \forall i | p'_i > p_i
\end{equation}
\noindent
In a final step, we bring the newly obtained and the old tests together 
\begin{equation}
t'_{ie} = \left \{
	\begin{array}{ll}
		0 & , p'_i \leq p_i \\
		t^*_{ie} - t_{ie} & , p'_i > p_i
	\end{array}
\right.
\end{equation}
to find the number of testing that needs to be done in addition.
\subsubsection{Minimizing Risk}
\noindent
We want to find the distribution of new tests for a given number \(m\), which minimizes risk when there are already tests. The idea is similar to the previous case. Improvements can only be achieved for bins with new \(p'_i\), which are more likely than old \(p_i\). Given $I' = \{i \mid p'_i \leq p_i\}$ as before, we compute new values $t^*_{ie}$ for $i \in I \setminus I'$:
\begin{equation}
\label{eq:ra-solution-minimize-R-t-update}
t^*_{ie} = \frac{\sqrt{\lambda_e \epsilon_e p'_i}(m + 2 \left|H\right|\left|I \setminus I'\right|)}{\left(\sum_{j \in H} \sqrt{\lambda_j \epsilon_j}\right) \left(\sum_{k \in I \setminus I'} \sqrt{p'_k}\right)} - 2 %, \forall p'_i > p_i
\end{equation}
Consequently, we distribute new tests only between these bins
\begin{equation}
t'_{ie} = \left \{
	\begin{array}{ll}	
		0 & , p'_i \leq p_i \\
		t^*_{ie} & , p'_i > p_i
	\end{array}
\right.
\end{equation}
and set all other to zero.

\noindent
% ============================================================================================
%
% ============================================================================================
\section{Evaluation}\label{sec:evaluation}
\begin{figure}
\centering
	\includegraphics[width=0.45\textwidth]{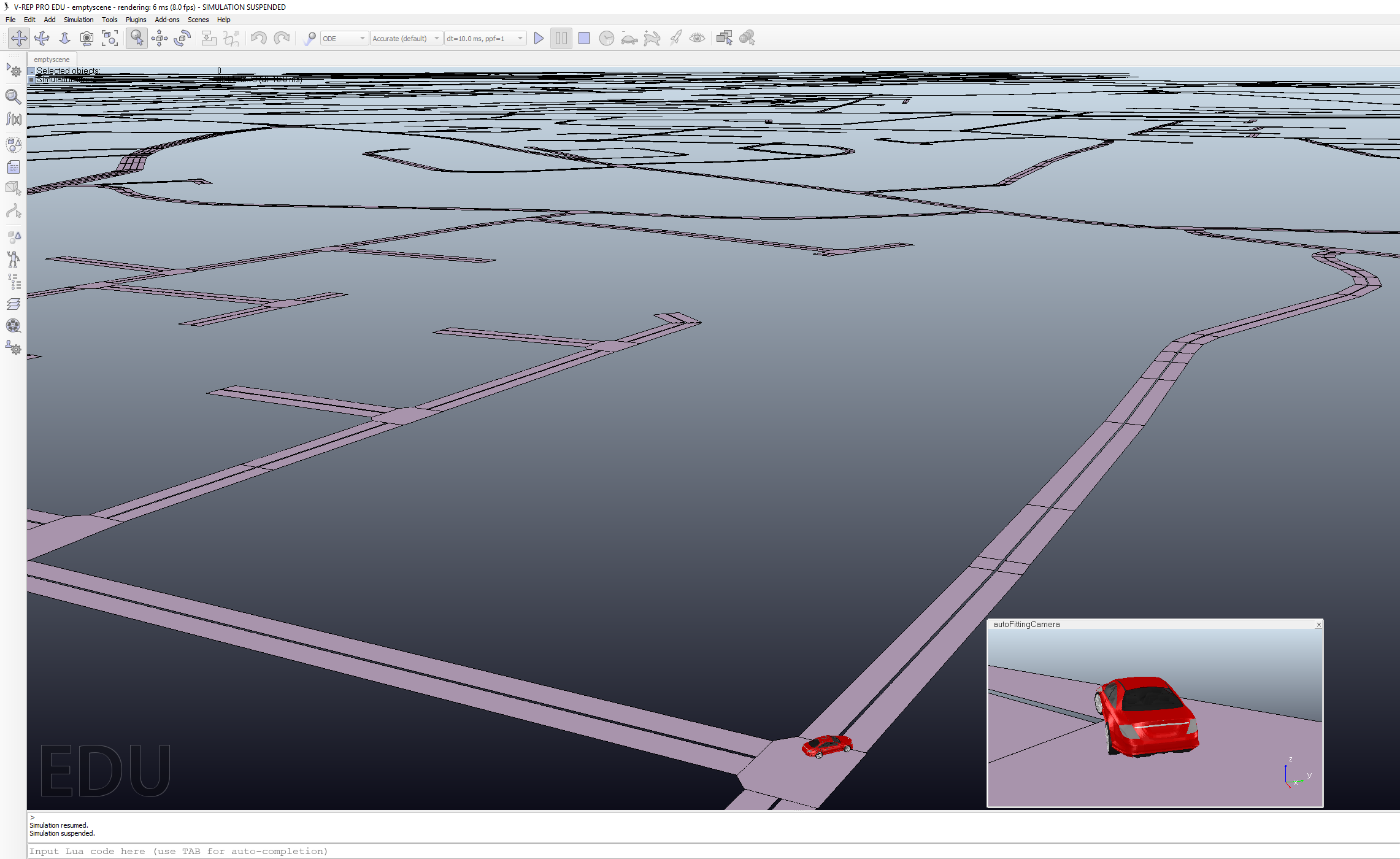}
	\caption{3d physics simulation of our scenario}
	\label{fig:3dphys}
\end{figure}

\noindent
For our evaluation, we implemented the scenario described in Section \ref{sec:examplearch} inside a 3D physics simulation environment (see Figure \ref{fig:3dphys} and appendix \ref{app:casestudy}). This enabled us to evaluate our approach without an actual fleet of vehicles. The simulation environment showed how risk is affected when our approach is not applied (uncontrolled risk with only deployment time tests) versus the application of the different strategies (controlled risk). 

\subsection{Setup}	
\noindent
We started our case study with an operational profile estimate based on one city. As a critical scenario, we assumed the aforementioned tire blowout. An estimate for \(\lambda\) is based on a field study in \cite{ratrout_evaluation_2011} that observed a tire blowout once in driving 16,278 kilometers and only because of an accident. Therefore, we assumed a \(\lambda_e \leq \frac{1}{16,278 * 200} \). The product of the inequality is the result of 200 planning steps (demands) per kilometer in our example. The severity \(\epsilon_e\) was set to 1. The subdomains where chosen as described in the appendix with 200 bins. We set the risk upper bound value to $1.0e-4$, which in a real setting is determined according to domain knowledge of specific safety requirement levels.

\begin{figure}
\centering
	\includegraphics[width=0.49\textwidth]{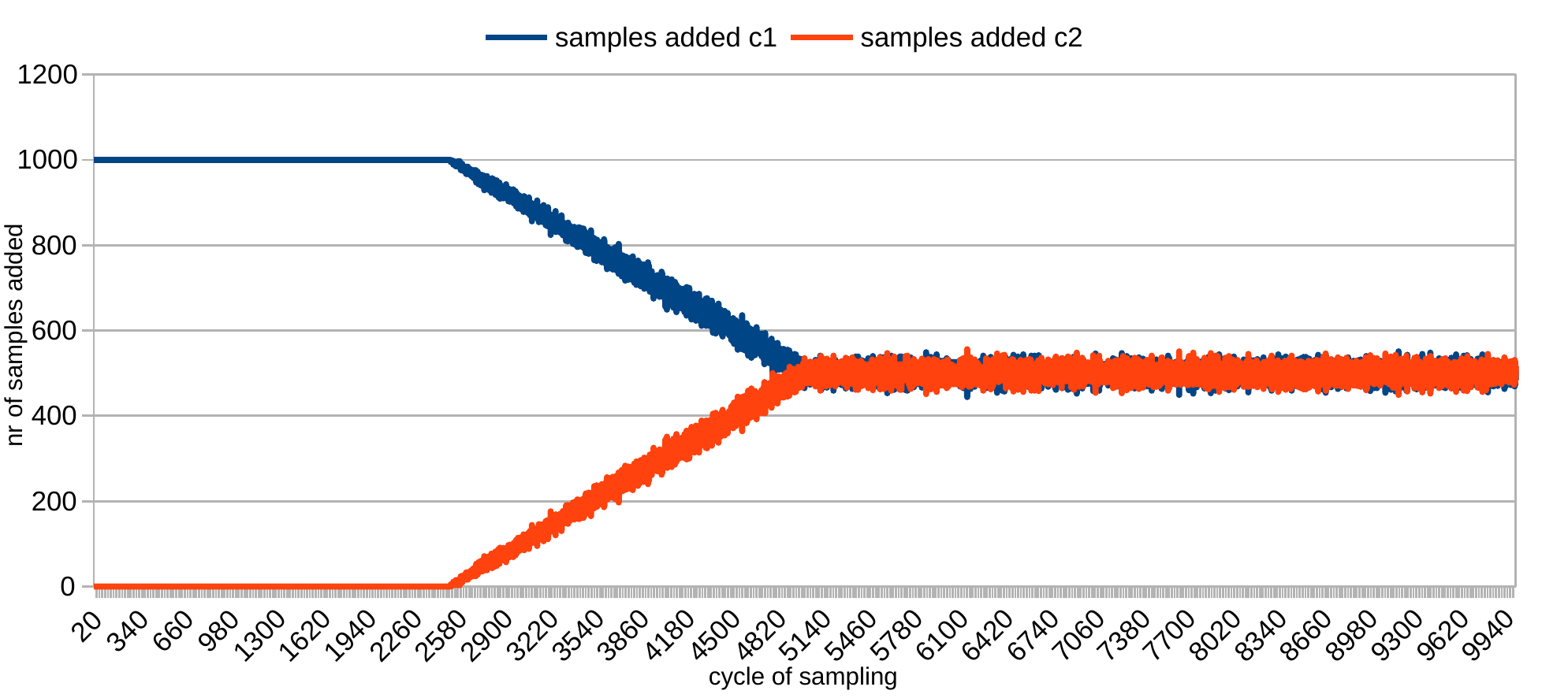}
	\caption{Profile sampling}
	\label{fig:profile}
\end{figure}

\begin{figure}
\centering
	\includegraphics[width=0.49\textwidth]{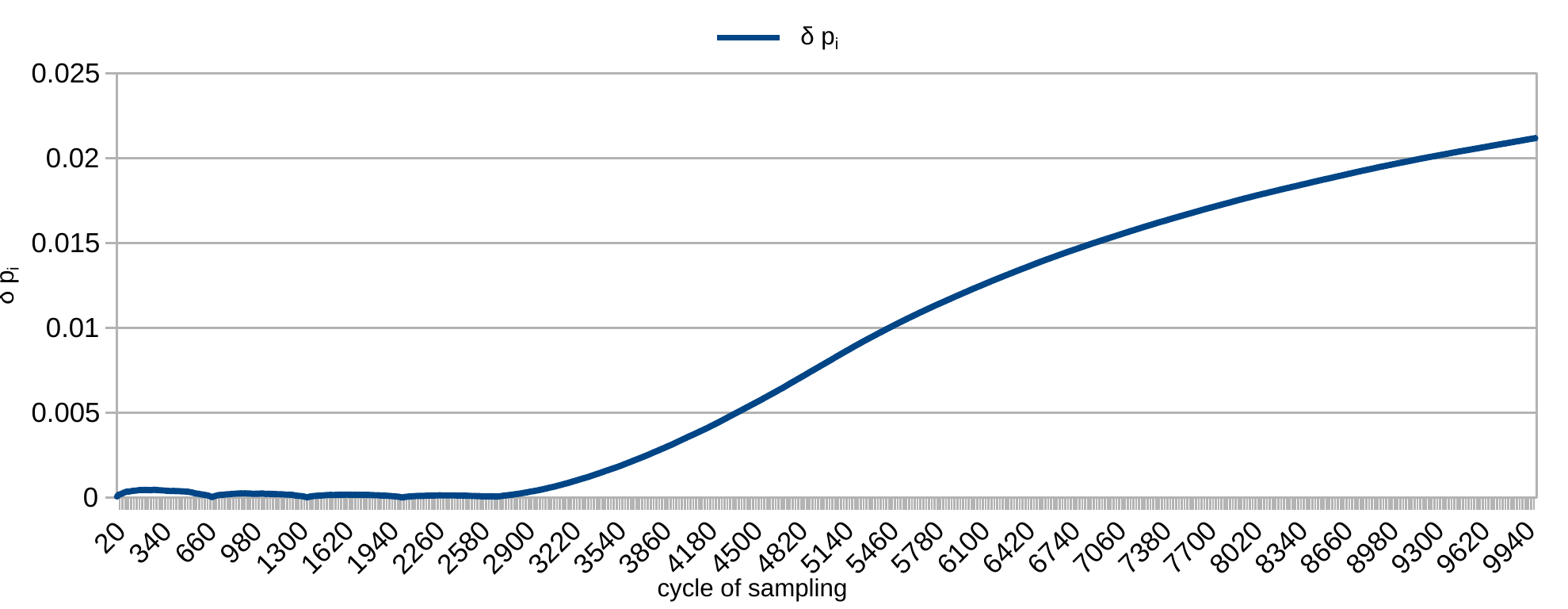}
	\caption{Delta of current profile to initial profile}
	\label{fig:deltaprofile}
\end{figure}

\noindent
We emulated our environment change by a transition of sampling from two different data sources (two cities), as shown in Figure \ref{fig:profile}. This resulted in a change in profile (approximated by the area between profiles) from the initial distribution over time, as depicted in Figure \ref{fig:deltaprofile}. The x-axis represents a sequence of cycles with a growing number of operational profile samples.

\noindent
\subsection{Ensuring an Upper Bound}
\begin{figure}
\centering
	\includegraphics[width=0.49\textwidth]{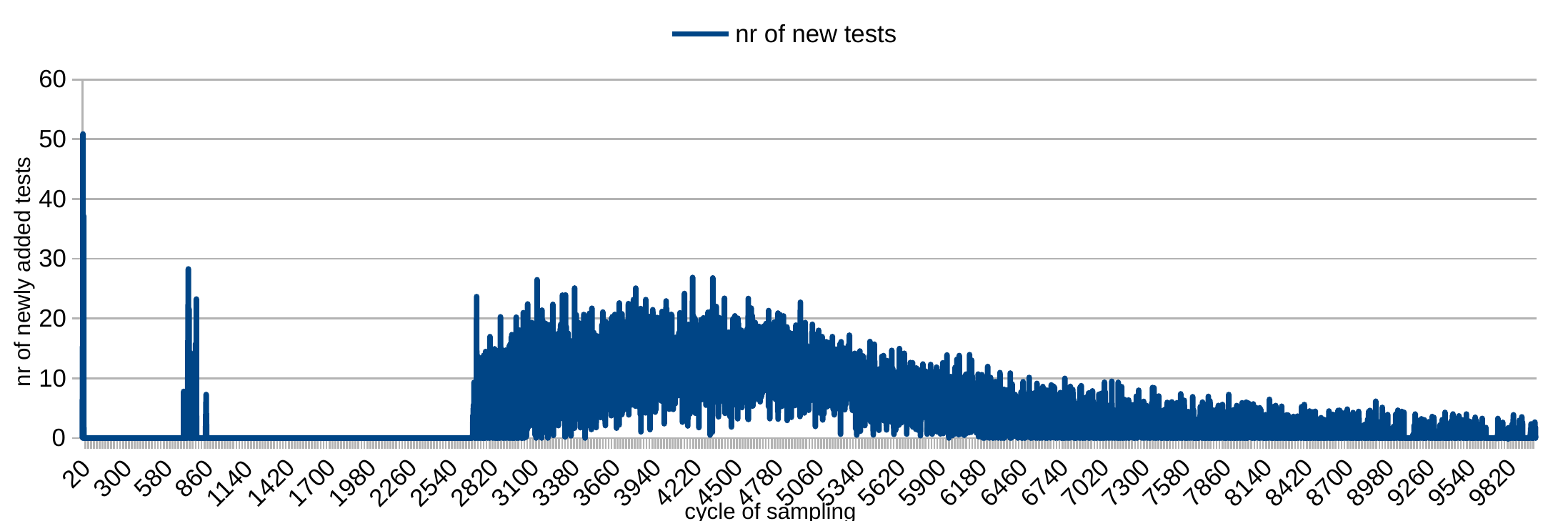}
	\caption{Numbers of added tests when maintaining the upper bound with strategy 1.}
	\label{fig:maintainbound-added}
\end{figure}

\begin{figure}
\centering
	\includegraphics[width=0.49\textwidth]{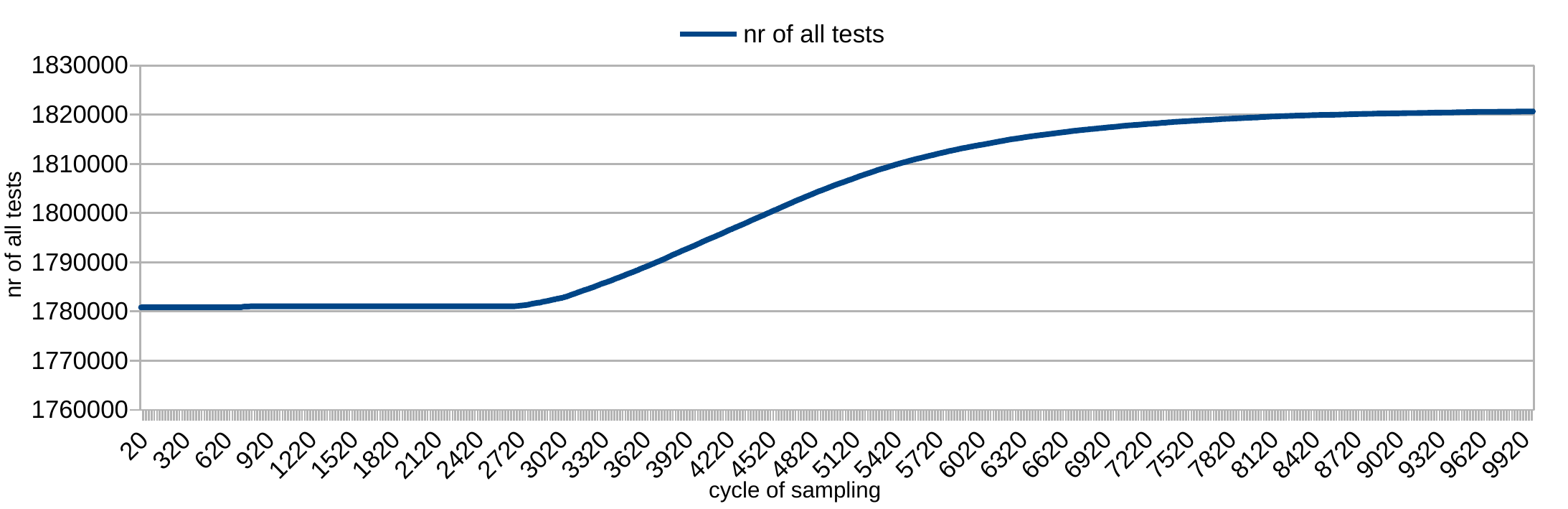}
	\caption{Total number of tests for maintaining the upper bound with strategy 1}
	\label{fig:maintainbound-total}
\end{figure}

\noindent
Based on the strategy for maintaining the upper bound, we were able to keep the upper bound below the required level (omitted because it is a simple line). The necessary tests that were allocated each cycle are shown in Figure \ref{fig:maintainbound-added}. As expected, for the small fluctuations in the profile during the deployment in only one city, almost no additional tests are necessary. But as soon as the sampling enforces a change in the profile, tests become necessary to mitigate the change regarding risk. In a later phase, when the profile is approaching a new steady-state, testing becomes less of a need again. In Figure \ref{fig:maintainbound-total} we see the total number of tests, which shows similarity in shape to the change in profile. 

\subsection{Continuous Addition of Tests}
\begin{figure}
\centering
	\includegraphics[width=0.49\textwidth]{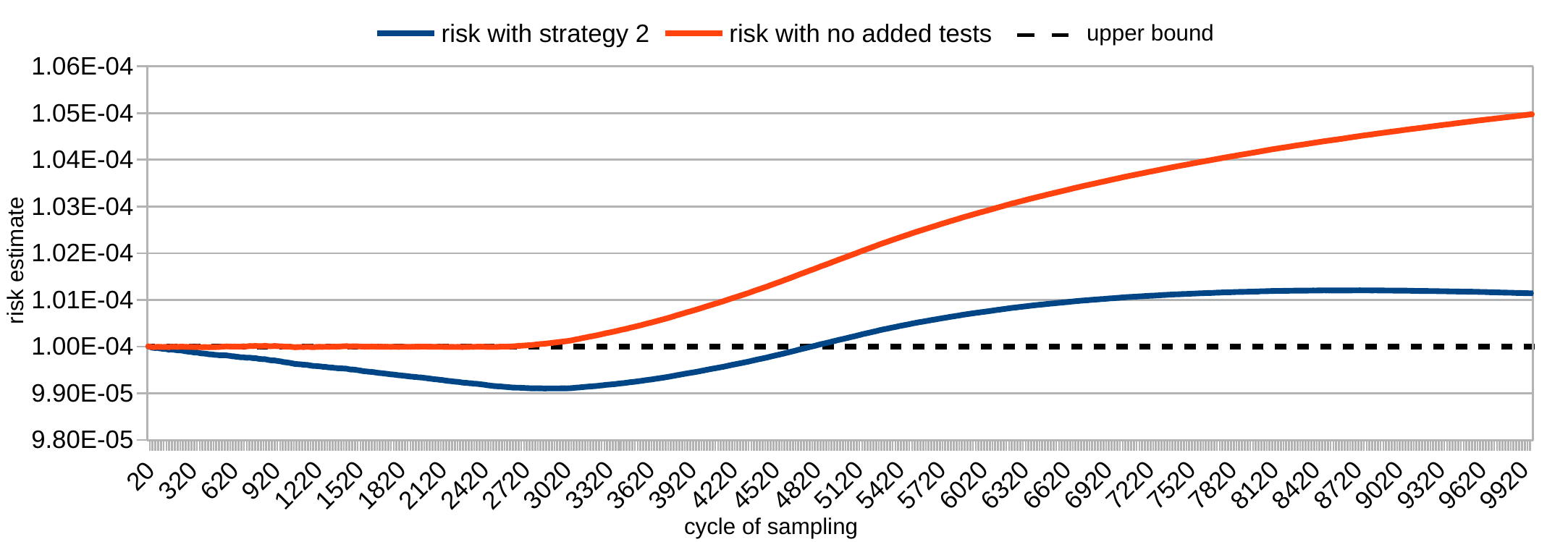}
	\caption{Risk for constantly adding tests with strategy 2 and risk without additional tests}
	\label{fig:adding-tests}
\end{figure}
\noindent
Risk predictions grow increasingly divergent when transitioning from the first to the second city (see Figure ~\ref{fig:adding-tests}). This divergence is solely originated by the uncontrolled risk (red line) as the controlled risk (blue line) first fluctuates below the risk upper-bound but finally breaks through. This is accomplished by a fraction of initial tests, i.e., adding 200 on top of the 1.8 million tests (0.01\%). 
While this would be expected for a safety-critical system that was thoroughly tested before the first deployment, these results suggest two interesting reflections: (1) how fine-grained the feedback control actuation is to keep the risk below the upper-bound and (2) how important is to have guidance on where to allocate these few tests among the various equivalence classes. 
\subsection{Continuous Addition of Tests while Ensuring the Upper Bound}
\begin{figure}
\centering
	\includegraphics[width=0.49\textwidth]{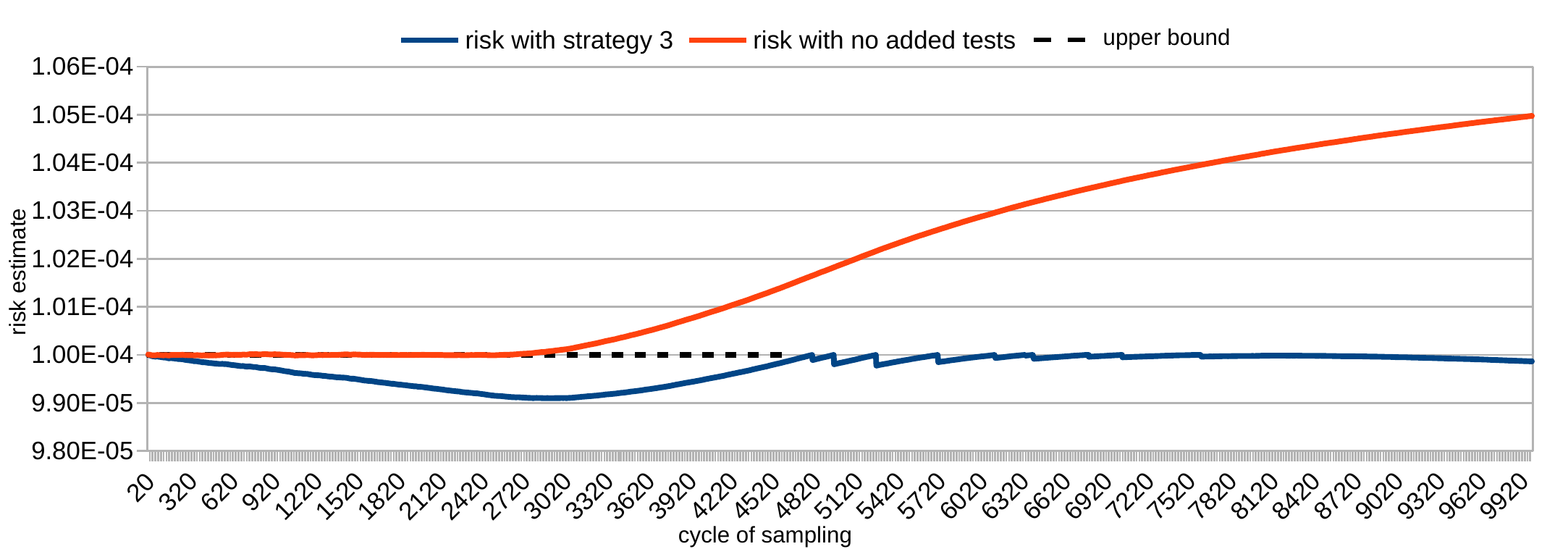}
	\caption{Risk if adding tests constantly while ensuring the upper bound with strategy 3 }
	\label{fig:add-and-maintain}
\end{figure}
\noindent
In the previous case, we could continue adding tests to keep the risk below the upper bound. Nonetheless, we still need a way to overcome the situations where the number of tests planned in a cycle might not be enough. 

We investigated this situation with the third strategy that combines the previous two strategies. The results of this combined strategy are shown in Figure \ref{fig:add-and-maintain}. Comparing the blue curves from Figure \ref{fig:add-and-maintain} with Figure \ref{fig:adding-tests}, we can see that strategy 3 provides two forms of improvements: (1) it kept the risk curve at a lower level than strategy 2 and (2) it corrected for risk more strongly for the cycles where the risk started to increase more quickly. These improvements stem from continuously adding tests (strategy 1) while still estimating the need for additional tests (strategy 2).

\noindent
Ultimately, our evaluation illustrated the difficulty of keeping an autonomous safety-critical system operating below a risk upper bound. The outcomes of the three strategies provide an intuition of this difficulty.

% ============================================================================================
%
% ============================================================================================
\section{Threats to validity}\label{sec:discussion}

\emph{External validity} discusses the situations for which the research assumptions and outcomes might not generalize to a different but relevant setting~\cite{wohlin2012experimentation}. One generalizability threat is how representative are differences between the testing and usage profiles. To mitigate this threat, we selected two cities with distinct street patterns that have an effect on the driving profile (speed, direction). Another relevant threat related to how representative the dataset is of real-world self-driving data. We mitigated this threat by designing a data generation process that can be parameterized to various particular self-driving situations. Although we used two distinct cities, the evaluation showed that even small topological differences already impose a challenge for test allocation. A third threat lies in the limitation on a fixed number of subdomains in a fixed input space. One might argue that this prevents us from dealing with unforeseen events because these would represent a new subdomain/bin where tests could be allocated to. We avoided this issue by trading state space size for data sparsity. The consequence is that the unforeseen events in our model correspond to bins with zero counts, i.e., almost zero probability of occurrence (as pointed out in Section \ref{subsec:approach-od@rt}). 
 
\emph{Internal validity} is the most common validity concern ~\cite{siegmund2015views}, and it evaluates if evidences of our experimental interventions were the necessary causes of the observed effects. 
We assumed that the data points (bins) are i.i.d., which might not always be true. The consequence is that in the worst-case scenario with dependent bins, the outcome would be a higher than expected risk measure, meaning more tests than we actually estimated for certain bins. We deemed the risk of non i.i.d. bins to be small, but we plan to address this situation in future work. Another validity situation is that any increase in the number of tests per equivalence class can only positively affect reliability. We avoided this threat by stating the assumption of equal probability of failure for each input within each class and that tests have no side-effects in the system. However, these internal validity assumptions also depend on the validity of the measurements~\cite{trochim2001research}, which we discuss next.

\emph{Construct validity} concerns the situations for which the operational indicators do not measure the actual concepts (constructs). This might happen through bias in the definitions, operations, and methods ~\cite{wieringa2014design} applied to the constructs. Two of our constructs are the most sensitive to biases: reliability and risk. The reason is that we do not measure them directly, instead we derived them from other directly measured constructs. We mitigated this threat by formalizing all the equations and procedures to compute reliability and risk. 

\emph{Conclusion validity} concerns the situations for which there are violations in the assumptions of the statistical methods that we adopted. The most relevant situations are the choices of the likelihood (data generation process) and the prior (density of tests). Wrong choices might bias the approximation of the posterior distribution of the risks over the equivalence classes. We mitigate this threat by relying on principles of statistical test methods~\cite{GardinerK1999} and taking a conservative approach by assuming an uninformative prior, e.g., the \(Beta(1,1)\) distribution.

% ============================================================================================
%
% ============================================================================================

\section{Related Work - Testing for Reliability}\label{sec:related}
\noindent
Testing for software reliability, as we presented in this paper, holds similarities with methods for testing self-adaptive systems and self-driving cars.

\subsection{Testing Self-Adaptive Systems - SAS}
\noindent
Testing SAS at run time is used to provide assurance that the system will behave as designed ~\cite{deLemos2017assurances}. This is particularly difficult for SAS because of the uncertainties about the impacts of system reconfigurations or environment changes~\cite{Giese:2014ca}, which are all inherent to the unanticipated operational scenarios~\cite{bertolino2019changing}. Hence, solutions for testing SAS focus on mitigating different types of uncertainty in the models of the system and the environment.

\emph{System model uncertainties} involve the types of failures that might happen and their impact at runtime. Online testing for SAS was proposed as a means to anticipate failures and trigger adaptation when corresponding tests fail~\cite{hielscher2008framework}. Although the failures could be a measure of reliability, online testing is still dependent on a complete knowledge of the operational scenario, which our approach precludes. Regarding uncertainties that lead to a degraded operation, Camara et al.~\cite{camara2014testing} investigated the resilience of SAS with respect to changes in the execution load or the partial failure of a system controller. Our approach is complementary as it allows these types of robustness tests to be allocated at runtime. 

\emph{Environment model uncertainty}. Reichstaller et al.~\cite{reichstaller2018risk} investigated a reinforcement learning approach to identify the priority of tests, which were modeled as policies (action state pairs) to maximize a given risk-based reward function. Besides the challenge in determining the reward function, their approach requires fine-grained state-level data that might not be available in the sparse operating environment of a SAS. Environmental uncertainties were also mitigated by adaptive testing techniques~\cite{fredericks2018empirical}. Our approach extends these techniques by adding reliability testing models.

Chen et al. \citep{chen2018design} mitigated both system and environmental uncertainties by means of a knowledge-base that formally specifies operational states, action sequences (trajectories), and corresponding constraints. We believe such a knowledge-base, particularly the operational trajectories, could be used as a prior in tailoring our clustering approach to a particular system-environment configuration.

\subsection{Testing Self-Driving Cars}
\noindent
The reliability and safety of autonomous vehicles are among the main topics listed by members of academia and industry~\cite{koopman2016challenges,knauss_2017}. Testing these systems involve novel types of uncertainties with respect to (1) internal behavior of systems with blackbox and stochastic machine learning models~\cite{koopman2016challenges}, (2) the lack of data on rare or low-frequency  events~\cite{koopman2018heavy}, and (3) missing or compromised data acquired online over large geographical regions~\cite{knauss_2017}. 

\emph{Testing Machine Learning Models}. Burton et al. ~\cite{burton2017making} proposed a notation-based approach to identify and mitigate uncertainties in the learned behavior of a machine learning model, e.g., insufficient training data, under-representative testing data, and difficulty to explain blackbox implementations. Machine learning models were also used to mitigate the uncertainty of the complex behavior that has to be learned. Wolf et al. \cite{wolf_adaptive_2018} investigated a reinforcement learning approach to learn the maneuver decisions while adopting a compact semantic state representation and ensuring a consistent model of the environment across scenarios.

Although the concerns of testing machine learning models seem orthogonal to reliability testing, we assume that the initial tests cover the critical behaviors of the autonomous vehicles. Otherwise reliability estimates would be compromised by a defected product.

\emph{Rare-Event Testing}. Low-Hutchinson et al.~\cite{hutchison2018robustness} developed a platform that mitigates the uncertainty of rare events by automatically generating tests. Their approach combines a data dictionary with safety invariant definitions and mutations of live data. Our approach is complementary in the sense that it could be integrated into their platform to prioritize test execution. Rare events were also obtained via simulation and used to test autonomous vehicles for scalability~\cite{o2018scalable}. Our approach can be complemented with more specific types of simulations, for instance, for stress testing the impact of catastrophic events as bridge collapses or flooded motorways.

\emph{Data Acquisition}. Regardless of the frequency of events, one still cannot ascertain when the new data of a fleet of autonomous vehicles will be available or even if the data could be trusted. This problem has been partially addressed by on-demand approaches like data synchronization methods~\cite{fritsch_time-bounded_2008} and the Tesla over-the-air updates, which is still vulnerable to data hacking~\citep{nie2018over}.

Ultimately, if we compare with the design of SAS, research on testing these systems is still lacking ~\cite{bertolino2019changing,siqueira2016characterisation}. As pointed out by Chechik et al. ~\cite{chechik2019software}, safety standards like DO-178C (aerospace) and ISO 262622 (automotive) provide recommendations on testing. However, they still lack the details of how to compose partial evidence of testing or how to use the results of one analysis to support the other. These are gaps that we expect that our work could help bridge.
% ============================================================================================
%
% ============================================================================================

\section{Conclusion and Future Work}\label{sec:conclusion}
\noindent
We developed a risk model based on the probability of occurrence and a corresponding hazard. The risk model combines a systematic identification of equivalence classes and the statistical testing of these classes. We built a simulation environment and a set of statistical models to compute the probability of distinct classes of hazardous scenarios (accidents). 

Our contributions were three-fold. A method for collecting and clustering multi-vehicle operational scenarios (driving trajectories) to mitigate the data sparsity. Statistical models that estimate the risk of accidents in a new traffic environment. A feedback control-loop model that actively minimizes the increase in the risk of accidents by allocating test appropriately.

The results were promising in a sense that we were able to measure and control risk for a diversity of operational scenarios obtained from two real cities with distinct street patterns. To allow the reproduction of our results, we made the procedures, models, and data publicly available to the community~\cite{github2020seams}.

Our future work will incorporate the uncertainties in the environment model and the runtime model~\cite{Giese:2014ca}, e.g., street repairs, time-of-day traffic changes (school pick up times), and distinct types of vehicles (taxis, trucks and delivery robots). This might require the combination of synthetic operational data with real and manipulated data, which in turn would allow us to evaluate which tactics can effectively reduce these uncertainties~\cite{moreno2018uncertainty}. On the epistemic perspective, we also plan to study a more principled methodology for partitioning the operational profile. This is necessary to sustain the evolution of the context (user expectations), which many times happens in reaction ~\cite{leggett_who_2018} to the adaptation goals and behaviors of the new autonomous system.

We would also like to extend our approach to cover different monitoring strategies when updating the operational profile. Because the age of profile data has a relation with how much it adds to the ''true'' current profile, our approach could benefit from strategies that take data age importance into account (like \cite{pietrantuono_towards_2019}). Another relevant issue in this context is the number of changes in the environment and from the adaptation layer.

\section{Acknowledgment}\label{sec:acknowledgment}
The authors would like to thank the Hasso-Plattner Institute and its Research School for the funding provided.

\appendix
\section{Appendix}\label{app:casestudy}
\noindent
To generate the self-driving data in a way that it is reproducible and realistic, we developed a simulation of a self-driving car that can be deployed in different cities. 
\noindent
\paragraph{Scenario Realization}
\noindent
In this simulation, a vehicle is driving in a static environment. Multiple vehicles are simulated sequentially. 
\begin{figure}
\centering
	\includegraphics[width=0.47\textwidth]{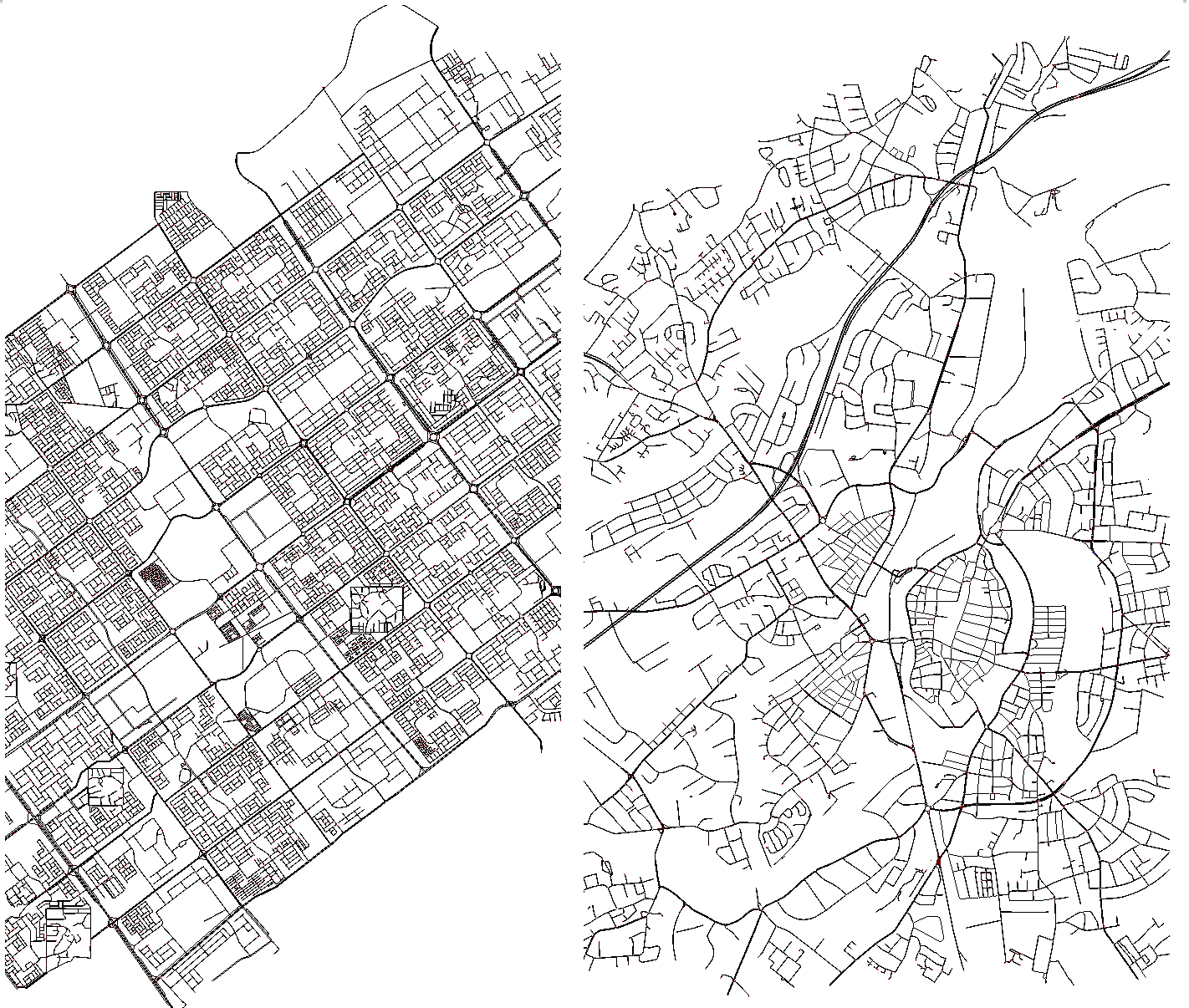}
	\caption{Scenario maps: Chandigarh and Luebeck}
	\label{fig:maps}
\end{figure}
\noindent
To simulate a changing environment, we used two areas that provide distinct characteristics with regard to their road layout (Luebeck in Germany and Chandigarh in India, see Figure \ref{fig:maps} and \cite{porta_alterations_2014}). We simulate the continuous change in the environment by combining different percentages of the data for a virtual collection window, from which we derive the operational distribution.  

\noindent
Regarding the adaptation engine, we designed it to continuously change the parameter set for the velocity planning according to the current environment. These parameters consist of maximal values for velocity, lateral acceleration, and longitudinal acceleration and de-acceleration. 
\paragraph{Implementation and Data Setup}
\noindent
The physical simulation is based on the V-Rep experimentation platform (see \cite{VREP2013}) together with the ODE-Simulator (see \cite{smith2003ode}). We based our definition of physical parameters of the vehicle on top of the standard Ackermann steering example shipped with the V-Rep platform. With respect to the control of the vehicle in the simulator, we implemented both the adaptation engine and the adaptable layer in Java language. The control part of the adaptable layer implements a pure pursuit control algorithm with variable look-ahead distance (see \cite{Paden2016}, Section V.A.1)). The areas are imported via SUMO from OpenStreetMap, resulting in navigable maps (\cite{behrisch_sumo_2011} \cite{OpenStreetMap}). The adaptation engine uses the map to compute a path via a simple Dijkstra shortest path implementation~\cite{dijkstra1959note} for a given target from the vehicle's current location.  

\noindent
Destinations for the vehicle are provided by a Java process, which also collects the data from the vehicles and stores them in a database. A list of predefined locations is randomly accessed and used as targets for the vehicle. While the vehicle is driving, it reports every planning horizon $\tau_t$ to the adapted layer and timely aligned to it an input from the environment $e_t = (pe_t, \vec{ve_t}, \vec{oe_t})$, with $t$ the time of collection. When storing $\tau_t$ and $e_t$ in the database, timestamps $t$ from the simulator are stored as well. 

\begin{figure}
\centering
	\includegraphics[width=0.3\textwidth]{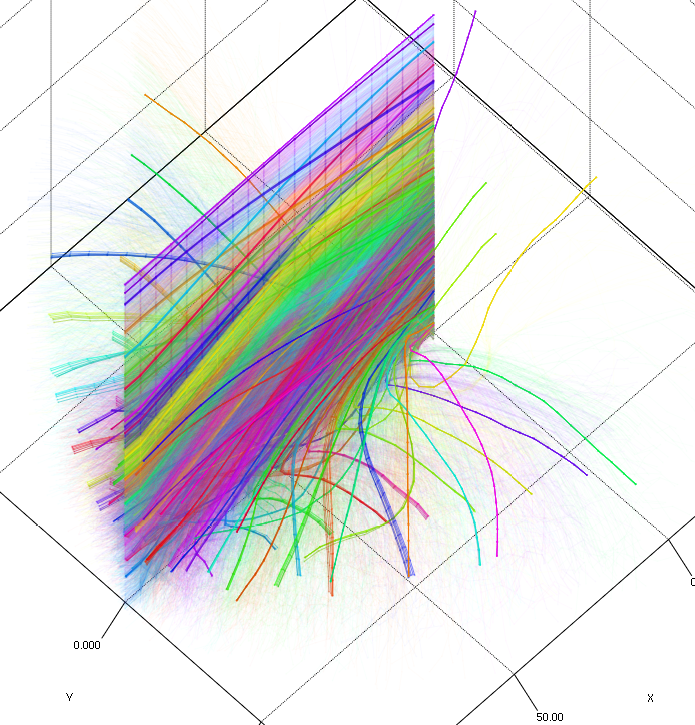}
	\caption{Visualization for the clustering with respect to \(\tau\) for part of the data}
	\label{fig:clustervis}
\end{figure}

\paragraph{Ground Truth} 
\noindent 
The vehicles drove more than 6334 kilometers in the simulated cities. Based on the collected $(\tau, e)$ data, we built a classification via k-means clustering (Figure \ref{fig:clustervis}). Counting the $(\tau, e)$ in the clustering provided the ground truth assumption for the ''true'' operational profile of the cities.

\paragraph{Environment Change Simulation by Sampling}
We derived a change in the environment by using data from two different maps. We collected ground truth profiles $p^{G_l}_i$ for a map of the city of Luebeck and $p^{G_c}_i$ for a map of the city of Chandigarh. In our scenario, vehicles are first deployed in the city of Luebeck and then in Chandigarh. An extension of the deployed vehicles environment from Luebeck to Chandigarh from an operational profile point of view is the average of $p^{G_l}_i$ and $p^{G_c}_i$. If $(\tau, e)$ are reported by the deployed vehicles at run-time, and a profile is incrementally built, we expect the overall monitored operational profile to be initially close to $p^{G_l}_i$ and after an infinite number of time to approximate the average of $p^{G_l}_i$ and $p^{G_c}_i$. We simulated this effect by incrementally sampling $(\tau, e)$ elements that were collected while building $p^{G_l}_i$ and $p^{G_c}_i$ in a changing ratio. More specifically, for a given time $t$, $p_i(t)$ is based on $n$ samples for which the likelihood of being from Luebeck is $(1-r(t))$ and $r(t)$ from Chandigarh with $r : T \rightarrow [0, 0.5]$ and $r(0) = 0$. 
\\

% ============================================================================================
%
% ============================================================================================
\bibliographystyle{ACM-Reference-Format}
\bibliography{lit}

\end{document}